\documentclass[
 reprint,
 amsmath,amssymb,
 aps,
]{revtex4-2}

\usepackage{graphicx}
\usepackage{dcolumn}
\usepackage{bm}
\usepackage{slashed}
\usepackage{hyperref}
\usepackage{subcaption}

\begin{document}

\preprint{APS/123-QED}

\title{Contribution of \texorpdfstring{\(\pi^0\)}{pi0} Exchange in Elastic Muon-Proton Scattering}

\author{Atharva Naik}
\email{anaik49@gwu.edu}
 \author{Andrei Afanasev}
 \email{afanas@gwu.edu}
\affiliation{
 Institute for Nuclear Studies, Department of Physics\\
 The George Washington University, Washington, DC, 20052, USA
}

\date{\today}

\begin{abstract}
The effect of the lepton's mass is significantly enhanced when the beam's energy is on the order of the lepton's mass. In the case of electrons, this corresponds to beam momenta on the order of a few MeV and is negligible in higher energy experiments.  In this study, we calculate the differential cross section \(d\sigma/ d\Omega\) for the helicity-flip meson exchange interference in elastic muon-proton \((\mu p)\) scattering \(\mu^- p \rightarrow \mu^- p\). In particular, we examine the \(\pi^0\) meson exchange in the \(t\)-channel for a longitudinally polarized beam and a transversely polarized target. We demonstrate the contribution to be larger for muons due to the lepton mass difference. Then we construct the corresponding beam-target double-spin asymmetries for the target polarized normal and parallel to the momentum transfer in the Breit frame, and then consider the model dependence of the calculation from the estimation of the \(\pi^0 \mu \mu\) vertex. The contribution was found to be on the order of \(\sim .15\%\) for muons in the kinematic region of the MUSE experiment.
\end{abstract}

\maketitle

\section{\label{sec:level1}Introduction}
The scattering of electrons on protons and neutrons has been used to probe the inner structure of hadrons for decades. Technological advancements in recent years have brought forth uncanny precision in the determination of observables in these experiments over a wide range of energies. Through scattering of electrons on protons, the determination of the ratio between the electric \((G_E)\) and magnetic \((G_M)\) proton form factors has been necessary in determining the inner electromagnetic structure of the proton. The two main experimental methods of extracting these Sachs Form Factors (SSFs) are the Rosenbluth separation technique and the polarization transfer method, and data has shown a significant discrepancy in these methods, which must be addressed \cite{Perdrisat:2006hj}.

The Rosenbluth separation method \cite{Rosenbluth:1950yq} relies on measuring the unpolarized differential cross section 
\begin{equation}
	\frac{d\sigma}{d\Omega} = C(\epsilon,Q^2) \left[ G_M^2(Q^2) +\frac{\epsilon}{\tau} G_E^2(Q^2)\right]\, ,
\end{equation}
where \(C(\epsilon,Q^2)\) is some kinematic factor that is well known, \(\epsilon = (1+2(1+\tau)\tan(\theta/2))^{-1}\) is a measure for the longitudinal polarization of the virtual photon, and \(Q^2 = -q^2 =-t\) is the momentum transfer squared. Then for a given \(Q^2\), measuring this cross section over a range of \(\epsilon\), gives the ratio of the form factors due to their linear dependence on \(\epsilon\). This method often serves as a test of the validity of the Born approximation in higher energy electron-proton scattering where the electron mass is neglected.

The polarization transfer method \cite{Akhiezer1974polarization, Arnold1981polarization}, providing another way to extract the ratio for the form factors \(G_E/G_M\), relies on measuring the polarization of the recoil proton parallel to its momentum \((P_l)\) and normal to its momentum \((P_t)\). By simultaneously measuring the degree of polarizations in each direction, the ratios of the form factors can be determined using
\begin{equation}
	\frac{P_t}{P_l} = - \sqrt{\frac{2\epsilon}{\tau(1+\epsilon)}}\frac{G_E}{G_M} \, .
\end{equation}
The discrepancy in the form factors from these two techniques has been one of the most surprising discoveries made at Jefferson Lab and the problem has been confirmed at higher and higher momentum transfers over the last two decades \cite{Perdrisat:2006hj}.

In 2010, the ultra precise measurement of the proton's charge radius of \( r_{ch} = .84184(67) \,\text{fm} \) \cite{Pohl:2010zza} and then \(r_{ch} =.84087(39) \, \text{fm} \) \cite{Antognini:2013txn} in 2013 via muonic hydrogen, were inconsistent with the previous averages by 5\(\sigma \) from \(ep\)-scattering data. Two methods of extracting the protons radius are from Lamb shift measurements in the hydrogen atom and from elastic \(ep\)-scattering. Inspired by this discrepancy, the problem has been named the ``proton radius puzzle" and motivated a series of new measurements at the Muon-Proton Scattering Experiment~(MUSE) at the Paul-Scherrer Institute in Switzerland as well as new \(ep\)-scattering measurements from the Proton Radius 2~(PRad-II) experiment at Jefferson Lab~(JLab) \cite{PRad:2020oor}. MUSE will utilize elastic \(\mu^\pm p\) and \( e^\pm p\)-scattering  simultaneously in hopes to resolve the puzzle \cite{Cline:2021ehf}. Operating at 115, 153 and 210~MeV/c beam momenta, a \(Q^2\) range of \(0.0016 -0.0799 ~\text{GeV}^2 \) for muons, and a scattering angle range of \( (20^\circ-100^\circ) \) \cite{MUSE:2017dod}, MUSE will be able to access a kinematic region where the muon mass cannot be neglected. Therefore we emphasize the importance of the helicity flip meson- and pseudoscalar-meson exchange amplitudes that are allowed only when the incoming lepton's masses are taken into account. 

In recent years, the variance in measurements from these techniques has brought a lot of attention to electromagnetic and hadronic corrections beyond the Born approximation, such as the two-photon exchange (TPE) effects. The challenges that these processes pose, lie in the need for modeling the intermediate nucleon's substructure which is model dependent and cannot be calculated without additional assumptions \cite{Afanasev:2017gsk}. Direct measurement of the TPE interference with the single-photon exchange provides information about the underlying TPE effects. The real (dispersive) part of the interference with TPE amplitude is odd in powers of \(e \). Therefore measuring the charge asymmetries accessible by MUSE provides direct way to evaluate TPE effects \cite{Afanasev:2020ejr}. Due to the large uncertainties in the hadronic contributions, its believed TPE effects might be the solution to the above problems.

On the other hand, the determination of the anomalous magnetic moment of the muon, \(a_\mu\), provides a stringent test of the modern theoretical framework for particle physics. The muon~\(g-2\), being one of the most precisely measured and theoretically well predicted quantities in physics, provides a way for us to probe new standard model physics by examining the discrepancy between measurement and theory. It was found that the hadronic contributions to \(a_\mu\) have the largest uncertainties. The Hadronic Light-by-Light (HLbL) contribution, arising in the rare scattering of a pair of photons to form a pseudoscalar meson which decays into another pair of photons, is being studied extensively as being one of the two hadronic effects limiting the standard model's precision of \(a_\mu \) \cite{Asmussen:2018oip}. The helicity-flip meson exchange examined in this paper may provide insight into those calculations.

A previous result \cite{Koshchii:2016muj} presented the interference of the scalar \(\sigma\) meson exchange to the unpolarized \(\mu p\) scattering amplitude and modeled the \(\sigma \mu\mu \) vertex via the coupling of the \(\sigma\) to two virtual photons. Then it used the vector meson dominance (VMD) model and assumed the \(\sigma\) couples each photon via a virtual \(\rho\) meson, to compute the charge asymmetry.

In this paper, we begin in Sec.~\ref{Born Cross Section} by calculating the Born cross section for longitudinally polarized leptons and demonstrate the \(\pi^0\)-exchange contributes for the case of a transversely polarized proton target, while retaining the lepton mass, but doesnot contribute to the unpolarized cross section in the first order correction of QED. Then in Sec.~\ref{Pi0 Contribution} we calculate the leading \(\pi^0\) meson contribution to the polarized \((\mu p)\) scattering cross section and then the corresponding beam-parallel single-spin asymmetries in Sec.~\ref{Asymmetries}. Finally, in Sec.~\ref{Model Dependence}, we examine the model dependence of the result on the effective \(\pi^0 \mu \mu\) vertex \((f_p)\) in the kinematic region of MUSE. Since muons are born longitudinally polarized from charged pion decay, the muons used in experiments serve as a test of these calculations, provided that a polarized hydrogen target is used.

\section{\label{Formalism} Elastic Muon-Proton Scattering Formalism }
For the following elastic scattering process 
\begin{equation}
	l(k_1) +p(p_1) \rightarrow l(k_2) +p(p_2) \, ,
\end{equation}
the lepton and proton vector currents that are used to calculate the leading order single photon exchange contribution are
\begin{subequations}\label{currents}
	\begin{equation}
	J_{\mu} ^{1\gamma} = \overline{U}(p_2) \left( \gamma_\mu G_M(Q^2) -\frac{(p_1 +p_2)_\mu}{2M}  F_2(Q^2) \right)U(p_1) \, ,
\end{equation}
\begin{equation}
	j_{\mu}^{1\gamma} = \overline{u}(k_2)\gamma_\mu u(k_1)\, .
\end{equation}
\end{subequations}
Here \(G_M(Q^2)= F_1(Q^2) + F_2(Q^2)\) is given by the Dirac and Pauli form factors, \(M\) is the mass of the proton, and \(u \) \((U)\) and  \(\overline{u}\) \((\overline{U})\) are bispinors of the lepton (proton). The scalar (pseudoscalar ) \(\sigma\) \((\pi^0) \) meson exchange process contributess in the next order of QED due to the meson's coupling to the lepton via two-photons (\(\pi^0 \rightarrow \gamma \gamma \, , \,\sigma \rightarrow \gamma \gamma   \)), which are given by the effective scalar (pseudoscalar) currents \cite{Koshchii:2016muj}
\begin{subequations}
\label{eq:pseudocurrents}
\begin{equation}
j^s = f_s \overline{u}(k_2)u(k_1) \:\: \quad j^\pi = f_p \overline{u}(k_2)\gamma_5 u(k_1)\label{subeq:1}
\end{equation}
\begin{equation}
J^s = g_s \overline{U}(p_2)U(p_1) \:\: \quad J^\pi = g_p \overline{U}(p_2)\gamma_5 U(p_1)\label{subeq:2}	
\end{equation}
\end{subequations}
The contribution of the \(\sigma\) meson is treated by Koschii \cite{Koshchii:2016muj}. It was also noted that \(\pi^0\)-exchange does not contribute if the scattering particles are unpolarized. Here we consider a possibility of polarization so we only examine the \(\pi^0\) meson. The parameters \(f_{s(p)} =f_{s(p)}(Q^2)\) and \(g_{s(p)}=g_{s(p)}(Q^2)\) describe the coupling of the corresponding mesons to the lepton and proton respectively.

\begin{figure}[b] 
\includegraphics[width = 1\linewidth]{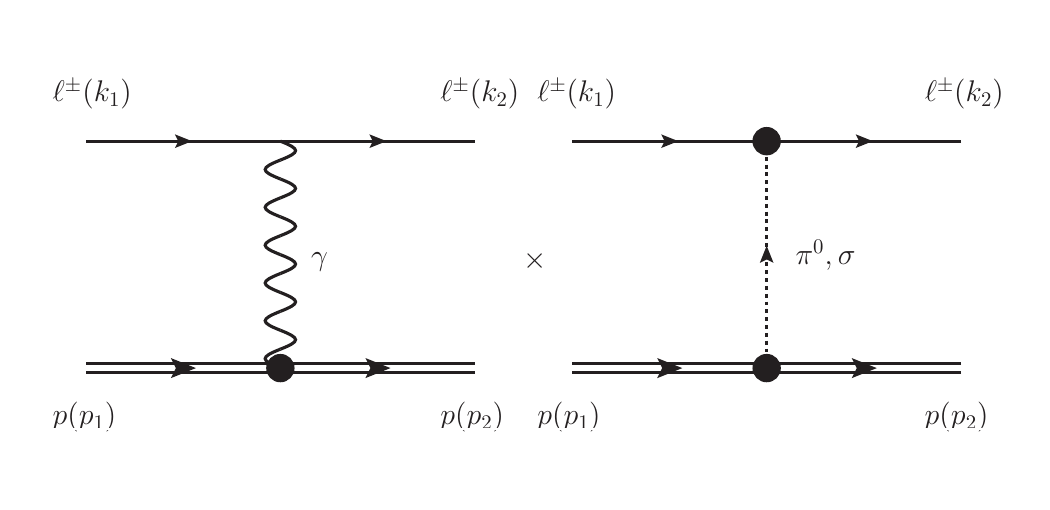}\caption{\label{fig:interference} Interference term between the single photon exchange and the single meson exchange diagrams in the \(t\)-channel.}
\end{figure}

The matrix elements for each of the diagrams in Fig.~\ref{fig:interference} for either lepton or antilepton \( (\ell ^\pm)\) is
 
\begin{subequations}
\begin{equation}\label{onephotoncurrent}
	\mathcal{M}_{1\gamma} = \mp \frac{ie^2}{Q^2} j_{\mu}^{v}J^{\mu v}
\end{equation}
\begin{equation}\label{onepioncurrent}
	\mathcal{M}^{\pi} = - \frac{i}{Q^2 +m_p^2} j^{\pi}J^{\pi}
\end{equation}
	\end{subequations}

The total square of the matrix element is 
\begin{align}
	|\mathcal{M}|^2  &= |\mathcal{M}_{1\gamma}+ \mathcal{M}_s +\mathcal{M}_{\pi}|^2\\
	&\approx|\mathcal{M}_{1\gamma}|^2 + 2\text{Re}\left[ \mathcal{M}_{1\gamma}\mathcal{M}_{\pi}^* \right] + 2\text{Re}\left[ \mathcal{M}_{1\gamma}\mathcal{M}_{s}^* \right].\label{eq:matrixelem}
\end{align}
	In the rest of the paper, we neglect the \( \mathcal{M}_{s}\) terms involving the \(\sigma\), since they can be added separately and also ignored \(|\mathcal{M}_{\pi}|^2\) since that is higher order in QED. Also since the \( f_p\) coupling is essentially the contribution arising from the axial anomaly diagram, \(f_p \sim  \alpha^2\).
	
We choose the Breit frame which is defined as the frame where no energy is transferred between the lepton and the proton so the initial and final nucleon/lepton energies are the same. We define  
\begin{eqnarray*}
	p_1 = (E_B, -\frac{\vec{q}_B}{2}), \quad \quad k_1 = (\varepsilon_B,\vec{k}_{1B})\\
	p_2 = (E_B, \frac{\vec{q}_B}{2}), \quad \quad k_2 = (\varepsilon_B, \vec{k}_{2B})
\end{eqnarray*}

To evaluate the energies and momenta, we use the usual convention choosing the \(z\)-axis parallel to the three-momentum transferred to the nucleon as the coordinate system. The exact expressions can be found in Appendix ~\ref{BreitAppendix}.

The energy of the lepton before and after the collision, retaining the lepton's mass, is 
\begin{equation}
	\varepsilon_B = \sqrt{m^2 + \frac{Q^2}{4\sin^2 \frac{\theta_B}{2}}}\, ,
\end{equation}
where \(\theta_B \) is the scattering angle of the lepton in the Breit frame. The energy of the nucleon is 
\begin{equation}
	E_{1B} = E_{2B} = M\sqrt{1+\tau} \, ,
\end{equation} 
with \(\tau = \frac{Q^2}{4M^2}>0\). Lastly, the polarization of the initial nucleon and the initial lepton is encoded in the spinor sums

\begin{subequations}

\begin{equation}
	\sum_s \overline{U}_s(p_1) U_s(p_1) = (\slashed{p}_1 +M)\frac{1}{2}(1+ \gamma_5 \slashed{\xi}_N)
\end{equation}
\begin{equation}
	\sum_r \overline{u}_r(k_1) u_r(k_1) = (\slashed{k}_1 +m)\frac{1}{2}(1+ \gamma_5 \lambda)
\end{equation}
\end{subequations}
where \(\xi_p = (0, n_x,n_y,n_z)\) is the polarization vector, and \(\lambda = \pm 1\) describes the helicity of the lepton. The transverse polarization is suppressed additionally by the mass of the lepton due to the helicity structure of QED and therefore the longitudinal polarization dominates.

\section{\label{Born Cross Section} Born Cross Section}

The matrix element squared for the single photon exchange is modified from \cite{Dombey:1969wk} to include the lepton mass in \(L_{\mu\nu}\), and include the polarizations in the other directions. The lepton tensor is

 \begin{align}
 	L^{\mu \nu}&=\frac{1}{2} \text{Tr}\left[(m + \slashed{k}_1)\gamma^\mu (m+ \slashed{k}_1)(1+\gamma_5 \lambda) \gamma^\nu \right] \\
 	& = 2\left(\frac{q^2}{2}g^{\mu \nu} + k_{2} ^{\mu} k_{1} ^{\nu} +k_{1}^{\mu} k_{2}^{\nu} +i\lambda \epsilon^{\mu\nu\alpha \beta}k_{1\alpha} k_{2\beta}\right) \,  ,
 \end{align}
 of which the calculation of each of the nonzero elements can be found in Appendix~\ref{SPEAppendix}.
 Now for the hadronic tensor, we use the two-component spinor formulation of the nucleon current using the density matrix for the nucleon \cite{Rekalo:2002gv}, and use the form which has the initial proton polarized also found in Appendix~\ref{SPEAppendix}. 
 
 Then the matrix element squared can be calculated in the Breit frame, found in Appendix~\ref{MatrixElementAppendix}, and after conversion to the lab frame we arrive at Eq. (\ref{MatrixElementLab}). Note from this expression, the target must be polarized in the lepton's scattering  plane (see Appendix~\ref{SPEAppendix}). Therefore only in the \(x\)- and \(z\)-directions which form the scattering plane. Essentially, this is because the Born amplitude has zero imaginary part since the proton current is Hermitian.
 
 From this, the formula for the Born differential cross section from the matrix element squared is given by \cite{Berestetskii:1982qgu} in Eq. (\ref{eq:borncrosssection})

\begin{widetext}
\begin{align}\label{MatrixElementLab}
	|\overline{\mathcal{M}}|^2_{1\gamma}  &=\frac{4M^2e^4}{Q^2}\Biggl[ \left(\frac{4m^2}{Q^2} G_E^2 + 2\tau G_M^2\right) +\frac{4|\vec k_1|^2|\vec k_2|^2 }{Q^4(1+\tau)} \sin^2 \theta\left(G_E^2 +\tau G_M^2 \right) \nonumber \\
	 & +\mp \frac{\lambda Q}{M^2}\sqrt{m^2 + \frac{Q^2}{4}\left(1+\frac{4|\vec k_1|^2|\vec k_2|^2 }{Q^4(1+\tau)} \sin^2 \theta \right)} G_M^2 n_z \mp \frac{\lambda Q}{M}\frac{2|\vec k_1||\vec k_2| }{Q^2\sqrt{1+\tau} } \sin \theta G_E G_M n_x \Biggr]
\end{align}
\begin{align}\label{eq:borncrosssection}
 	\frac{d\sigma_{1\gamma}}{d\Omega} &=  \frac{Q^2}{(4\varepsilon_1\varepsilon_2 -Q^2)}\Biggl[ \left(\frac{4m^2}{Q^2} G_E^2 + 2\tau G_M^2\right) +\frac{4|\vec k_1|^2|\vec k_2|^2 }{Q^4(1+\tau)} \sin^2 \theta\left(G_E^2 +\tau G_M^2 \right)+ \nonumber \\ 
 	 & \mp \frac{\lambda Q}{M^2}\sqrt{m^2 + \frac{Q^2}{4}+\frac{|\vec k_1|^2|\vec k_2|^2 }{Q^2(1+\tau)} \sin^2 \theta } G_M^2 n_z \mp \frac{\lambda Q}{M}\frac{2|\vec k_1||\vec k_2| }{Q^2\sqrt{1+\tau} } \sin \theta G_E G_M n_x \Biggr]\frac{d\sigma_M}{d\Omega}
 \end{align}
 \end{widetext}
with \(d\sigma_M / d\Omega\) being the Mott cross section for scattering off of a point-like object given by
\begin{equation}
	\frac{d\sigma_M}{d\Omega} = \frac{\alpha^2}{Q^4} \frac{M(4\varepsilon_1\varepsilon_2 -Q^2)\vec k_2^2 }{|\vec k_1|((M+\varepsilon_1)|\vec k_2| - \varepsilon_2 |\vec k_1 |\cos \theta)} \, ,
\end{equation}
and the fine structure constant \(\alpha = e^2/4\pi \approx 1/137\). To get Eq. (\ref{eq:borncrosssection}) in terms of only the incoming beam energy and the scattering angle in the lab frame, the momentum transfer \(Q^2\) can be written as 
\begin{equation}
	Q^2 = 2(\varepsilon_1 \varepsilon_2 -|\vec k_1| |\vec k_2 |\cos \theta -m^2)\, ,
\end{equation}
and the outgoing beam energy as \cite{Gakh:2014zva}

\begin{equation}
	\varepsilon_2 = \frac{(\varepsilon_1+M)(\varepsilon_1M+m^2) + \vec k_1^2 \cos \theta \sqrt{M^2 -m^2\sin^2\theta}}{(\varepsilon_1+M)^2 -\vec k_1^2\cos^2\theta}\, .
\end{equation}

Finally, one can write the differential cross section in the Born approximation as a sum of an unpolarized part and a polarized part. 
\begin{equation}
	\frac{d\sigma}{d\Omega} = \frac{d\sigma_0}{d\Omega} + \frac{d\sigma_{p}}{d\Omega} \, .
\end{equation}
In terms of \(Q^2 \) and a parameter \( \epsilon_m \), which represents the virtual photon's polarization in the ultra relativistic limit, the unpolarized Born cross section is and can be compared with \cite{Koshchii:2016muj}
\begin{equation}\label{eq:unpolBorn}
	\frac{d\sigma_0}{d\Omega} = \frac{1}{\epsilon_m(1+\tau)}\Biggl[ \tau G_M^2  + \epsilon_m G_E^2 \Biggr]\frac{d\sigma_M}{d\Omega} \, ,
\end{equation}
and the polarized part is

\begin{align}
	\frac{d\sigma_{p}}{d\Omega} &= \mp \frac{\lambda Q^2}{4M^2\tilde{\epsilon}_m (1+ \tau)}\left[ Q\sqrt{\frac{Q^2}{4} +\tilde{\epsilon}_m}\,G_M^2 n_z  + \right. \nonumber \\
	&\left. 2M\sqrt{\tilde{\epsilon}_m -m^2} G_M G_E n_x \right]\frac{d\sigma_M}{d\Omega}\, .
\end{align}
with 
\begin{equation}
	\tilde{\epsilon}_m = \frac{(2m^2-Q^2)}{2} \frac{\epsilon_m}{\epsilon_m-1} \, ,
\end{equation} 
and
\begin{equation}\label{eq:epsilon}
	\epsilon_m = 1 - 2(1+ \tau)\frac{2m^2 -Q^2}{4\varepsilon_1\varepsilon_2 -Q^2}\, .
\end{equation}
Here \( \tilde{\epsilon}_m\) is some variable chosen to simplify the expression.
Comparing the coefficients of \(G_E^2 \) in Eq.~(\ref{eq:borncrosssection}) and Eq.~(\ref{eq:unpolBorn}) and then using Eq.~(\ref{eq:angleconversion}) \((4\varepsilon_1\varepsilon_2 -Q^2) \) can be rewritten as
\begin{equation}
	4\varepsilon_1\varepsilon_2 -Q^2 = \frac{1+\tau}{1}\left[ \frac{4|\vec k_1|^2|\vec k_2|^2 }{Q^2(1+\tau)} \sin^2 \theta +4m^2\right] \, .
\end{equation}

\section{\label{Pi0 Contribution} Contribution of \texorpdfstring{\(\pi^0\)}{pi0} Exchange}
The neutral pion is the lightest of the \(\pi \) mesons and all the pseudoscalar mesons, therefore the lightest discovered in nature. The residual strong force, responsible for the binding of protons and neutrons in atoms, is due to pion exchange between the nucleons.

The interference between the single photon exchange diagram and the single pion exchange can be described by the second term in Eq.~(\ref{eq:matrixelem}) and can be computing using Eq.~(\ref{currents}) and Eq.~(\ref{eq:pseudocurrents}). The evaluation of the traces and the calculation of every term can be found in Appendix~\ref{InterferenceAppendix}. The total interference term as a function of the scattering angle and the incoming beam momenta in the lab frame is then 
\begin{align}\label{eq:interferencelab}
	2\text{Re}[\overline{\mathcal{M}_{1\gamma}\mathcal{M}_{1\pi}^*}&]  = \pm\frac{ 16e^2\lambda m  M\sqrt{1+\tau} \text{Re}[f_p^*g_p^*]}{Q(Q^2 + m_\pi ^2)}\nonumber \\
	&\left( \sqrt{m^2 + \frac{Q^2}{4}+ \frac{|\vec k_1|^2 |\vec k_2|^2\sin^2\theta}{Q^2(1+\tau)}} \, F_1 n_z \right. \nonumber \\
	& + \left. \frac{4M|\vec k_1||\vec k_2|\sin \theta}{Q^2} G_M n_x \right) \, .
\end{align}
As follows from the Appendix~\ref{InterferenceAppendix} derivation, the single-spin target asymmetry due to normal polarization (\(y\)-direction) does not arise because of the pseudoscalar nature of the \(\pi^0\)-exchange.
Note this expression is proportional to the \(\pi^0\)'s coupling to the leptons as well as the mass and the polarization of the incoming lepton. This interference being proportional to the mass, also means the contribution will be significantly enhanced for muons. In the case of massless or unpolarized leptons, this term vanishes.

\section{\label{Asymmetries}Beam-target Double-Spin Asymmetries}
In this section, we address the  beam-target double-spin asymmetry associated with a transversely polarized target and a polarized lepton beam with positive or negative helicity, all in the lab frame. The asymmetry for the target polarized in the \(x\)-direction is defined as
\begin{equation}
	A_x = \frac{d\sigma_x^+ - d\sigma_x^-}{d\sigma_x^+ + d\sigma_x^-} \, ,
\end{equation}
where \(d\sigma_x^+ \) \( (d\sigma_x^-)\) denotes the differential cross section with \(n_z =0 \) and positive (negative) beam helicity, \(\lambda =+1 \) \((\lambda = -1)\). Similarly for the target polarized in the \(z\)-direction, we set \(n_x=0 \), and compute the same asymmetry. We found

\begin{align}
	A_x &= \mp \frac{2|\vec k_1| |\vec k_2|\sin \theta}{M Q \sqrt{1+\tau}} \Biggl[ 2 G_E G_M  \nonumber \\
	&- \frac{4mg_p \text{Re}[f_p^*](1+\tau)G_M}{\alpha \pi  (Q^2 +m_\pi^2)}\Biggr] /\frac{d\sigma_{R}}{d\Omega}\, ,
\end{align}
and 
\begin{align}
	A_{z} &= \mp\frac{Q}{M^2}\sqrt{m^2 + \frac{Q^2}{4}+\frac{|\vec{k_1}|^2|\vec{k_2}|^2 }{Q^2(1+\tau)} \sin^2 \theta }\Biggl[ G_M^2 \nonumber \\  
 	&-\frac{ mM g_p \text{Re}[f_p^*] \sqrt{1+\tau}F_1 }{\alpha\pi (Q^2+m_\pi^2)}  \Biggr]/\frac{d\sigma_{R}}{d\Omega}\, ,
\end{align}
where \(d\sigma_R /d\Omega \) is the reduced Born cross section defined by
\begin{align}
	\frac{d\sigma_R}{d\Omega} = \frac{d\sigma_0}{d\Omega}/ \frac{d\sigma_M}{d\Omega} &= \left(\frac{4m^2}{Q^2} G_E^2 + 2\tau G_M^2\right) \nonumber \\
	&+\frac{4|\vec{k_1}|^2|\vec{k_2}|^2 }{Q^4(1+\tau)} \sin^2 \theta\left(G_E^2 +\tau G_M^2 \right) \, .
\end{align}

The first term in the asymmetry can be seen as the born approximation to the asymmetry, and the second term as the correction to the asymmetry due to the interference with the pion exchange.
The asymmetries in terms of \(\epsilon_m \) and \(Q^2\) can also be found using the relations in Sec.~\ref{Born Cross Section}

\begin{align}
	A_x &= \mp \frac{ 4 \sqrt{\tilde{\epsilon_m} -m^2}}{M} \Biggl[ G_E G_M  \nonumber \\ 
	& - \frac{2mg_p \text{Re}[f_p^*](1+\tau)G_M}{\alpha \pi (Q^2 +m_\pi ^2)}\Biggr] /\frac{d\sigma_{R}}{d\Omega}\, , 
\end{align}

\begin{align}
	A_z &= \mp \frac{Q}{M^2}\sqrt{\frac{Q^2}{4} + \tilde{\epsilon}_m}\Biggl[ G_M^2 \\
	& -\frac{ mM g_p \text{Re}[f_p^*] \sqrt{1+\tau}F_1 }{\alpha\pi (Q^2+m_\pi^2)}  \Biggr] / \frac{d\sigma_R}{d\Omega}\, ,
\end{align}

 The relative size of the corrections compared to the Born approximations are

\begin{align}
	\frac{\delta A_x}{ A_x} &= \frac{2m M g_p \text{Re}[f_p^*](1+\tau)}{\alpha \pi(Q^2+m_\pi^2) G_E}\, ,\\
	\frac{\delta A_z}{A_z} &= \frac{mMg_p\text{Re}[f_p^*]\sqrt{1+\tau} F_1}{\alpha \pi(Q^2 +m_\pi^2) G_M^2}\, .
\end{align}

\section{\label{Model Dependence}Model Dependence}

Currently, there is little to no experimental information about the \(\pi^0\mu \mu\) vertex, \(f_p \). The leading HLbL contributions to the muon \(g-2\) provide some theoretical knowledge about this interaction using an off-shell \(\pi^0 \gamma\gamma \) transition form factor, which is extensively studied using various low energy hadron models \cite{Jegerlehner:2009ry}. However, they often consider small mixing with the \( \eta\) and \( \eta' \) mesons.

  To approximate the \(\pi^0 \mu\mu \) form factor we assume, due to lepton universality, that the effective coupling should be the same as the \(\pi^0 e e \) coupling, which we estimate from the \(\pi^0 \rightarrow e^+ e^- \) decay width. This decay proceeds, to lowest order, in one loop by exchanging two virtual photons. Its suppressed by two orders of \(\alpha \) and approximate helicity conservation relative to the \(\pi^0 \rightarrow \gamma \gamma \) process \cite{KTeV:2006pwx}.

The decay width, to first order coupling, is given by \cite{Bergstrom:1982zq}
\begin{equation}
	\Gamma(\pi^0 \rightarrow e^+e^-) = \frac{f_p^2 m_\pi}{8 \pi}\sqrt{1-\frac{4m^2}{m_\pi^2}}\, .
\end{equation}
Then the coupling \(f_p \) is 
\begin{equation}\label{coupling}
	f_p = \left( \frac{8\pi \Gamma_i }{\sqrt{m_\pi^2 -4m_e ^2}} \right)^{1/2} \, .
\end{equation}
Using values obtained from the PDG \cite{ParticleDataGroup:2020ssz}, \(f_p = (3.08\pm .08)\times 10^{-7} \).
The dependencies of the proton Sachs form factors, \(G_E \) and \(G_M \) on \(Q^2 \) are experimentally well described up to \( \sim 6\, \text{GeV}^2\) using the Kelly parametrization \cite{Kelly:2004hm}
\begin{equation}
	G(Q^2) \propto \frac{\sum_{k=0}^{n}a_k \tau^k}{1+\sum_{k=1}^{n+2}b_k \tau^k} \, ,
\end{equation}
where \(a_k\) and \(b_k\) are parameters fitted through data.  We use a value for the \(\pi^0 pp \) coupling \(g_p \) given by Klomp \cite{Klomp:1991vz} and then convert the pseudovector pion coupling constant to the pseudoscalar pion coupling \cite{NavarroPerez:2016eli}. The margin of error associated with \(g_p \) is negligible for the purposes of this calculation. 

\begin{figure*}[t]
\centering
\begin{minipage}[b]{.49\textwidth}
\includegraphics[width =1\linewidth]{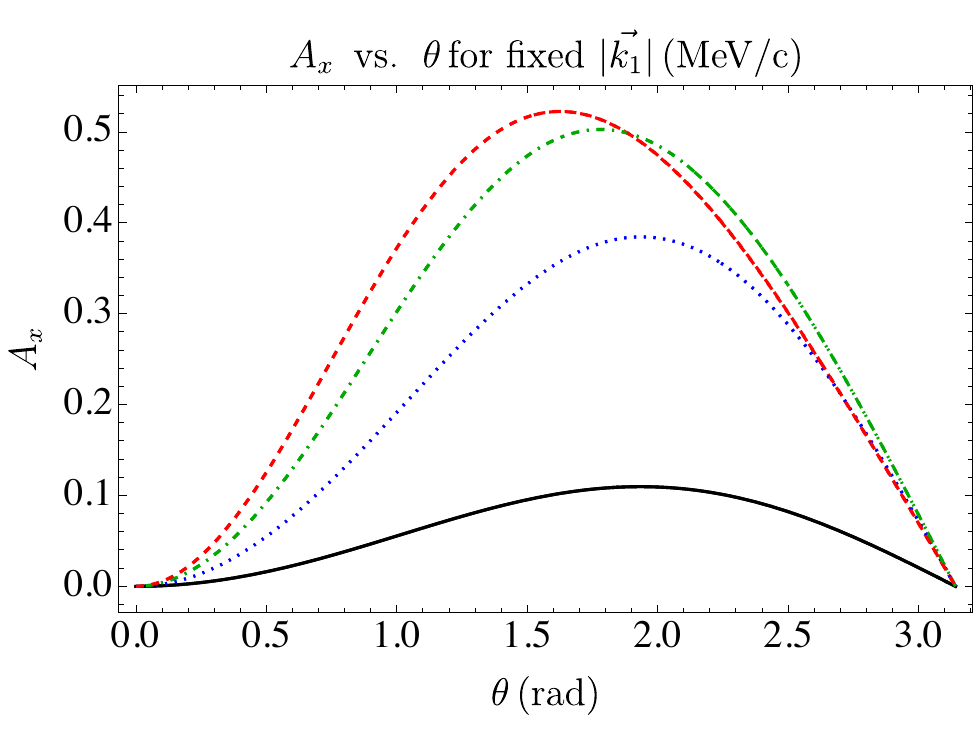}
\caption{Total beam-target double-spin asymmetry in the \(x\)-direction as a function of the scattering angle \(\theta \) for muon momenta \(|{\bf{k_1}}| = 100 \)\(\, \text{MeV/c}\) (black, solid), \(|{\bf{k_1}}| = 200 \)\(\, \text{MeV/c}\) (blue, dotted), \(|{\bf{k_1}}| = 300 \)\(\, \text{MeV/c}\) (green, dot-dashed), and  \(|{\bf{k_1}}| = 400 \)\(\, \text{MeV/c}\) (red, dashed).}\label{fig:Ax}
\end{minipage}\hfill
\begin{minipage}[b]{.49\textwidth}
\includegraphics[width =1\linewidth]{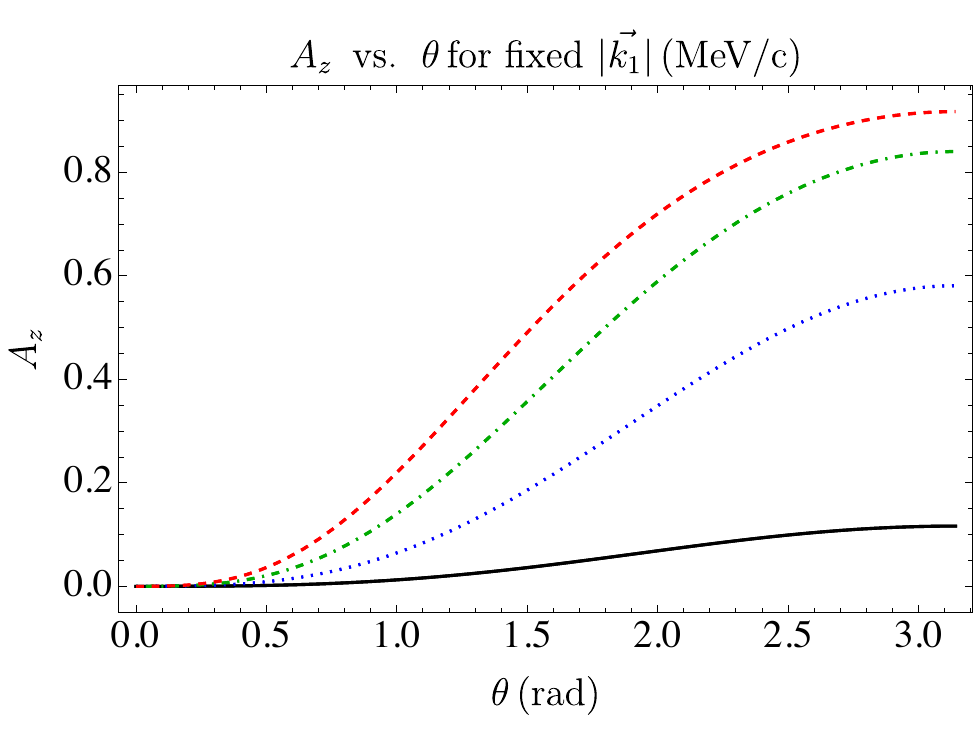}
\caption{Total beam-target double-spin asymmetry in the \(z\)-direction as a function of the scattering angle \(\theta \) for muon momenta \(|{\bf{k_1}}| = 100 \)\(\, \text{MeV/c}\) (black, solid), \(|{\bf{k_1}}| = 200 \)\(\, \text{MeV/c}\) (blue, dotted), \(|{\bf{k_1}}| = 300 \)\(\, \text{MeV/c}\) (green, dot-dashed), and  \(|{\bf{k_1}}| = 400 \)\(\, \text{MeV/c}\) (red, dashed).}\label{fig:Az}
\end{minipage}
\end{figure*}

\begin{figure*}
\centering
\begin{minipage}[b]{.49\textwidth}
\includegraphics[width =1\linewidth]{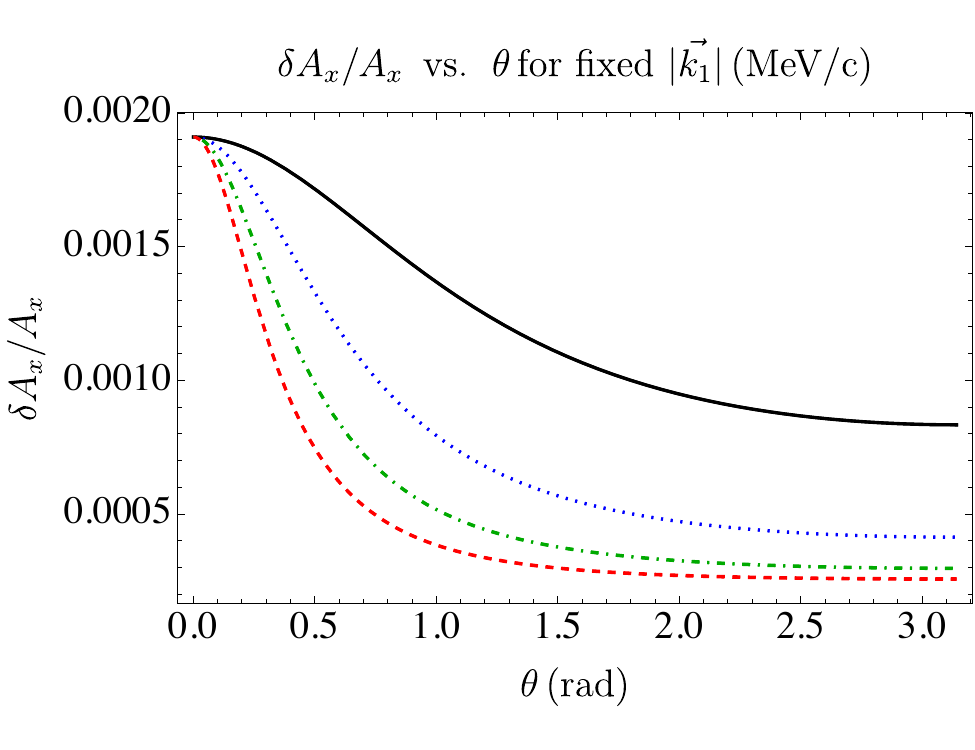}
\caption{Absolute size of the correction of the asymmetry relative to the Born asymmetry in the \(x\)-direction for muon momenta \(|{\bf{k_1}}| = 100 \)\(\, \text{MeV/c}\) (black, solid), \(|{\bf{k_1}}| = 200 \)\(\, \text{MeV/c}\) (blue, dotted), \(|{\bf{k_1}}| = 300 \)\(\, \text{MeV/c}\) (green, dot-dashed), and \(|{\bf{k_1}}| = 400 \)\(\, \text{MeV/c}\) (red, dashed).}\label{fig:delx}
\end{minipage}\hfill
\begin{minipage}[b]{.49\textwidth}
\includegraphics[width =1\linewidth]{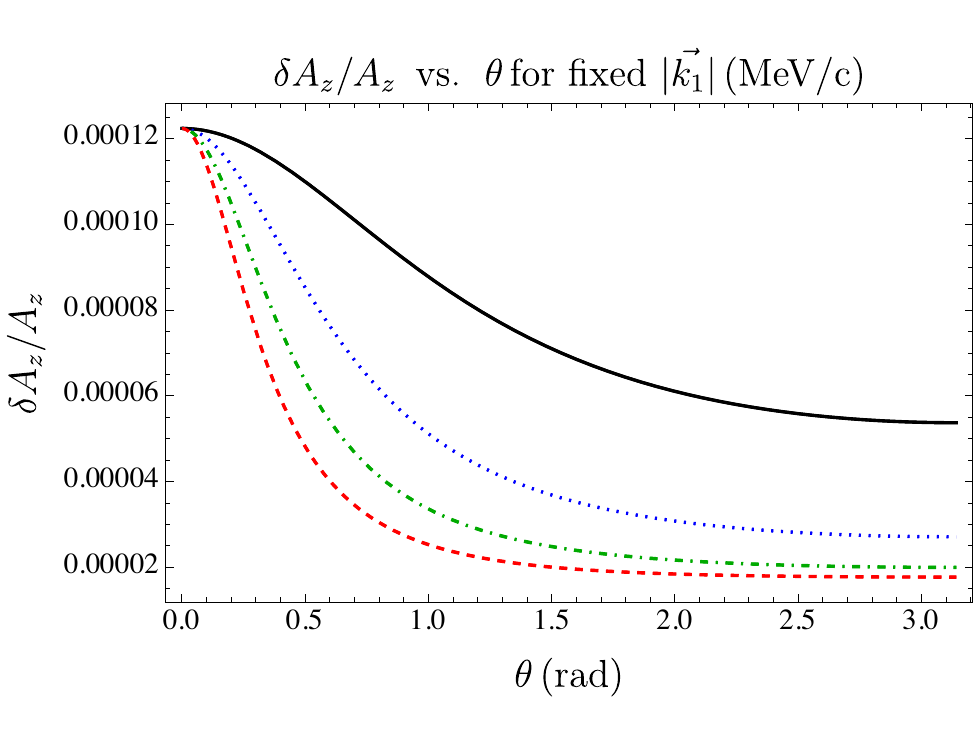}
\caption{Absolute size of the correction of the asymmetry relative to the Born asymmetry in the \(z\)-direction for muon momenta \(|{\bf{k_1}}| = 100 \)\(\, \text{MeV/c}\) (black, solid), \(|{\bf{k_1}}| = 200 \)\(\, \text{MeV/c}\) (blue, dotted), \(|{\bf{k_1}}| = 300 \)\(\, \text{MeV/c}\) (green, dot-dashed), and \(|{\bf{k_1}}| = 400 \)\(\, \text{MeV/c}\) (red, dashed).}\label{fig:delz}
\end{minipage}
\end{figure*}

The asymmetries as functions of the the beam momenta and scattering angle are provided in Figs. \ref{fig:Ax},\ref{fig:Az} using this constant \(f_p \).Then the size of the correction to the asymmetry due to the interference with the single pion exchange is compared to the Born asymmetry and plotted in Figs. \ref{fig:delx}, \ref{fig:delz} for momenta in the kinematic region explored by MUSE. All the figures include the upper and lower bounds associated with the Kelly parameters, but are practically invisible.

\begin{figure*}
\centering
\begin{minipage}[b]{.49\textwidth}
\includegraphics[width =1\linewidth]{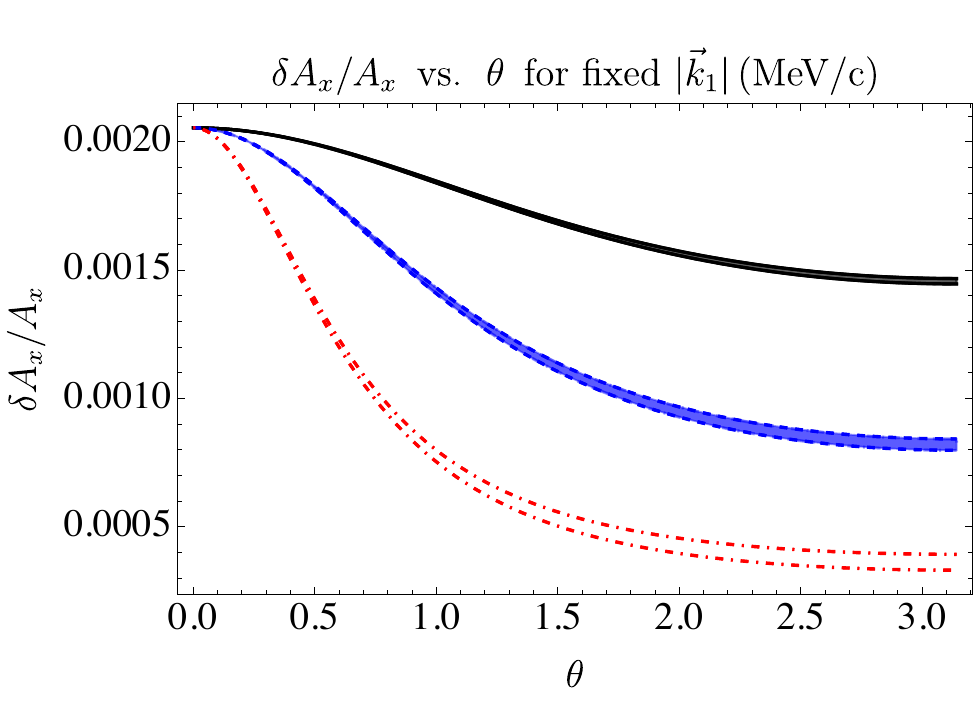}
\caption{Absolute size of the correction of the asymmetry, relative to the Born asymmetry in the \(x\)-direction for \( \Lambda^2 = .4 \div 1.6\)\(\, \text{GeV}^2/ \text{c}^2 \)  and momenta \(|{\bf{k_1}}| = 50 \)\(\, \text{MeV/c}\) (dark shaded region, black, solid lines), \(|{\bf{k_1}}| = 100 \)\(\, \text{MeV/c}\) (blue shaded region, blue, dashed lines) and \(|{\bf{k_1}}| = 200\)\( \, \text{MeV/c}\) (transparent region, red, dot-dashed lines).}\label{fig:modeldepx}
\end{minipage}\hfill
\begin{minipage}[b]{.49\textwidth}
\includegraphics[width =1\linewidth]{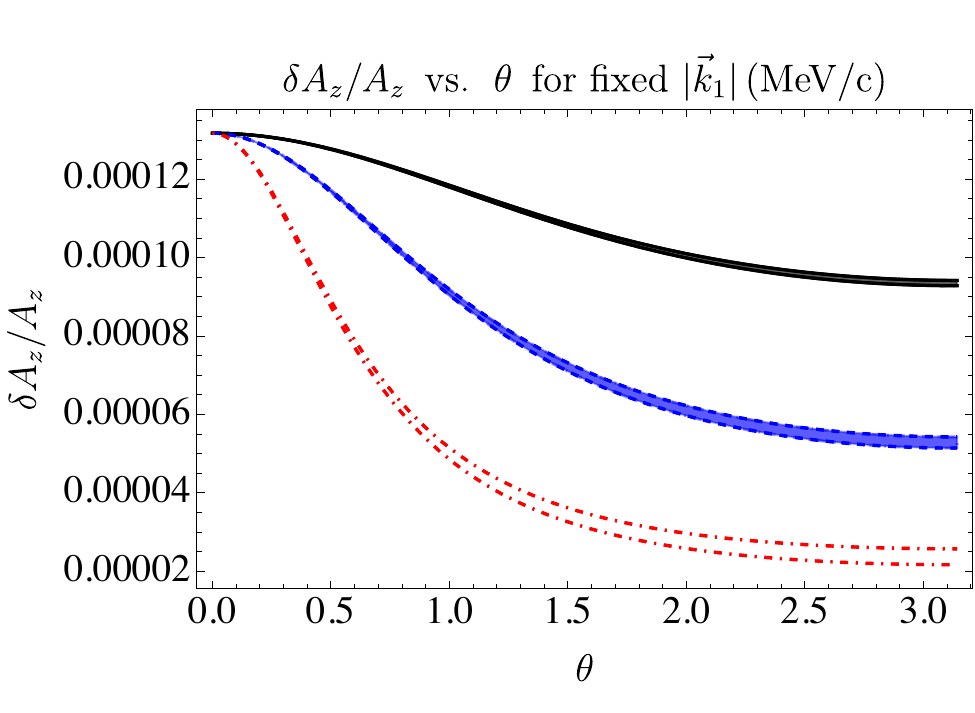}
\caption{Absolute size of the correction of the asymmetry, relative to the Born asymmetry in the \(z\)-direction for \( \Lambda^2 = .4 \div 1.6\)\(\, \text{GeV}^2/ \text{c}^2 \) and momenta \(|{\bf{k_1}}| = 50 \)\(\, \text{MeV/c}\) (dark shaded region, black, solid lines), \(|{\bf{k_1}}| = 100 \)\(\, \text{MeV/c}\) (blue shaded region, blue, dashed lines) and \(|{\bf{k_1}}| = 200\)\( \, \text{MeV/c}\) (transparent region, red, dot-dashed lines).}\label{fig:modeldepz}
\end{minipage}
\end{figure*}

For \( f_p\)'s dependence on \(Q^2 \), we model it using a monopole-like form factor
\begin{equation}
	f_p(Q^2) = \frac{f_{p}}{1+ \frac{Q^2}{\Lambda^2}} \, ,
\end{equation}
and estimate a range of values of \(\Lambda^2 \) based on the scalar radius of the pion. The scalar radius of the pion is \(\sim.6-.7\,\text{fm}\), so we chose a range of \(\Lambda^2 \in [ .4, 1.6]\, \text{GeV}^2 \) such that \(r \sim \sqrt{1/ \Lambda^2} \in [ 0.5, 1.0]\, \text{fm} \). The shaded and transparent regions in Figs. \ref{fig:modeldepx}, \ref{fig:modeldepz} show the variation in the correction resulting from varying the coupling constant that are the on the level of \( 10^{-4}\).

\section{\label{Conclusion}Conclusion and Discussion}
In this paper, we calculated the contribution of the pseudoscalar \(\pi^0 \) meson exchange's interference in polarized elastic lepton-proton scattering. We had to recompute the Born cross section for the process taking into account the initial particles polarizations and retaining the lepton's mass, which is commonly discarded when doing calculations for ultrarelativistic electrons. We only examined longitudinally polarized leptons since transversely polarized leptons are helicity suppressed by an additional factor of the lepton mass. We found, most importantly, that the size of the correction is on the order of \(\sim .15 \% \) for muons. Moreover the correction is \(\sim 10^2 \) times smaller in the \(z\)-direction than the \(x\)-direction and as \( Q^2 \rightarrow 0\), \(\delta A_x/\delta A_z \rightarrow 2\mu^2 \). 

When considering the dependence of the form factor on \( Q^2 \), the variation around the pion's scalar radius shows little deviation, and is negligible at small scattering angles. However higher beam momenta give considerably larger corrections, due to \( Q^2 \)\,'s dependence on \(|{\bf{k_1}}|\). 

One of the most crucial assumptions in this paper is assuming that lepton universality implies the \( \pi^0 \mu \mu\) coupling should be the same as \( \pi^0 ee\) and that the leading order contribution to it is mass independent. This may not be the case if the one loop exchange of two photons depends on the lepton's mass.
Given a sub-percent accuracy in current experiments to the proton's charge radius and TPE, the contribution of \(\pi^0\)-exchange is not negligible. Such a measurement can be realized for a polarized lepton beam and a polarized proton target. The obtained results should be considered an essential, and previously omitted, addition to the published calculations  \cite{Koshchii:2016muj, Tomalak2014two, Tomalak2018dispersion} of the hadronic conributions into TPE of muon-proton scattering.

\begin{acknowledgments}
We acknowledge the support of US National Science Foundation through Grant PHY-2111063. The Feynman diagrams were drawn on JaxoDraw-2.1-0, and the plots created in Mathematica 13.1.0.0.

\end{acknowledgments}

\appendix

\section{\label{BreitAppendix} Breit Frame Formalism}
The definition of the Breit frame is the frame in which there is no energy transferred between the leptons and nucleons. Thus the energy of the virtual particle vanishes, \(q_B~=~(0,\vec{q}_B)\) and 
\begin{eqnarray}
	\vec{p}_1 = (0,0,-\frac{q_B}{2})\, \quad \quad \vec{k}_1 = \frac{q_B}{2}(\cot\frac{\theta_B}{2}, 0 , 1)\, \\
	\vec{p}_2 = (0,0,\frac{q_B}{2})\, \quad \quad \vec{k}_2 = \frac{q_B}{2}(\cot\frac{\theta_B}{2}, 0 , -1)\,
\end{eqnarray}
and \(q_B = \sqrt{-q^2} = \sqrt{Q^2} \equiv Q \) .

\section{\label{SPEAppendix}Leptonic and Hadronic Tensors for Polarized Single Photon Exchange}

Using the form of the leptonic tensor
\begin{equation}
	L^{\mu\nu}= 2\left(\frac{q^2}{2	}g^{\mu \nu} + k_{2} ^{\mu} k_{1} ^{\nu} +k_{1}^{\mu} k_{2}^{\nu} +i\lambda \epsilon^{\mu\nu\alpha \beta}k_{1\alpha} k_{2\beta}\right)\, ,
\end{equation}
all the nonzero terms are
 \begin{align}
 	L^{00} &= 2\left( 2m^2+\frac{Q^2}{2}\cot^2\frac{\theta_B}{2}\right)\\
 	L^{11} &= \frac{1}{2}Q^2\left( 2+\cot^2\frac{\theta_B}{2}\right) \\
 	L^{22} &= Q^2\\
 	L^{12}&=-L^{21} = -2i\lambda \sqrt{Q^2}\sqrt{m^2 + \frac{Q^2}{4\sin^2 \frac{\theta_B}{2} }}\\
 	L^{10} &= L^{01} = 2Q \cot\frac{\theta_B}{2}\sqrt{m^2 + \frac{Q^2}{4\sin^2 \frac{\theta_B}{2}}}\\
 	L^{20}&= - L^{02} = -i\lambda Q^2 \cot\frac{\theta_B}{2}
 \end{align}

For the hadronic tensor we use the simpler form of the nucleon current \(\mathcal{J_\mu} = \chi_2^\dagger F_\mu \chi_1\), with \(F_\mu = 2mG_E\), for \(\mu =0\) and \(F_\mu = i\vec{\sigma}\times {\bf{q_B}} G_M\) for \(\mu = x,y,z\) given by \cite{Rekalo:2002gv}. Then the hadronic tensor can be written as 
\begin{equation}
	W_{\mu \nu}  = \frac{1}{2} \text{Tr}\left[ F_\mu \rho_1 F_\nu \rho_2 \right]\, ,
\end{equation}
 with \(\rho_1\)  \((\rho_2)\) being the \(2\times 2\) density matrix for an initial (final) nucleon with polarization \(\bf{P_1}\) \((\bf{P_2})\) given by \(\rho_{1,2} = \frac{1}{2} (1+ \vec{\sigma}\cdot \bf{P_{1,2}})\). For an initial polarized nucleon is easily written as 
 
 \begin{align}
 	 	W_{\mu\nu} &= W_{\mu\nu}^{(0)} + W_{\mu\nu}(\bf{P_1})\\
	 	& = \frac{1}{2} \text{Tr} \left[ F_\mu  F_\nu \right]+ \frac{1}{4} \text{Tr} \left[ F_\mu (1+ \vec{\sigma}\cdot {\bf{P_1}}) F_\nu\right]  \, . 
 \end{align}
Then the only non-zero terms are
 
\begin{align}
	W^{00}&= 4M^2 G_E^2\\
	W^{11} &= W^{22} = Q^2 G_M^2\\
	W^{10} &= -W^{01}\mp 2 Q MG_E G_M n_y\\
	W^{12} &=-W^{21}= \mp iQ^2 G_M^2 n_z\\
	W^{20} &= -W^{02} = \mp i Q MG_E G_M n_x
\end{align}

\section{\label{MatrixElementAppendix}Matrix Element for Single Photon Exchange}

	The matrix element squared in the Breit frame is simply calculated using 
	\begin{equation}
		|\overline{\mathcal{M}}|^2_{1\gamma B} = \frac{4M^2 e^4}{Q^2} L^{\mu \nu}W_{\mu\nu} \, ,
	\end{equation}

\begin{widetext}
	\begin{align}\label{MatrixElementBreit}
|\overline{\mathcal{M}}|^2_{1\gamma B}& =\frac{4M^2e^4}{Q^2} \left[ \left(\frac{4m^2}{Q^2} G_E^2 + 2\tau G_M^2\right) +\cot^2 \frac{\theta_B}{2}\left(G_E^2 +\tau G_M^2 \right) \mp \frac{\lambda Q}{M^2}\sqrt{m^2 + \frac{Q^2\csc^2\frac{\theta_B}{2}}{4} }G_M^2 n_z \mp \frac{\lambda Q\cot \frac{\theta_B}{2}}{M} G_E G_M n_x \right]\, 
\end{align} 
\end{widetext}
and the conversion to the lab frame is given by.
\begin{equation}\label{eq:angleconversion}
	\cot^2\frac{\theta_B}{2} = \frac{4|\vec k_1|^2|\vec k_2|^2 }{Q^4(1+\tau)} \sin^2 \theta \, .
\end{equation}

\section{\label{InterferenceAppendix}Lepton and Hadronic Tensors for the \texorpdfstring{\(\pi^0\)}{pi0} Interference term}

The interference term  can be computing using Eqs. (\ref{onephotoncurrent}) and (\ref{onepioncurrent})

\begin{align}
	2\text{Re}[\overline{M_{1\gamma}M_{1\pi}^*}] &= 2\text{Re} \left[\mp \frac{ie^2}{Q^2}j^v_\mu J^v _\mu \frac{+i}{Q^2 +m_\pi^2} (j^p J^p )^*    \right] \\ 
		& =\frac{\pm 2e^2}{Q^2(Q^2 +m_{\pi}^2)}\text{Re} [f_p^* g_p^* L_\mu W^\mu]\, , \label{interference formula}
\end{align}
where the Lepton tensor \(L^{\mu\nu}\), and the Hadron tensor \(W_{\mu\nu}\) are
\begin{align}
	L_\mu & = Tr[(m+\slashed{k}_2)\gamma_\mu (m+\slashed{k}_2)\frac{1}{2}(1+ \gamma_5 \lambda)\gamma_5]\\
	W^\mu &= Tr[(M+\slashed{p}_2)\Gamma^\mu (M+\slashed{p}_2)\frac{1}{2}(1+\gamma_5 \slashed{\xi}_N)\gamma_5 ]
\end{align}
Evaluating the traces
\begin{align}
	L_\mu &= 2 \lambda m (k_1+k_2)_\mu \\
	W^\mu &=  2(p_2 \cdot \xi_p)((F_2 -2G_M)p_1^\mu +F_2 p_2^\mu) \\
		& \quad+4G_M(p_1 \cdot p_2 -M^2)\xi_p ^\mu 
\end{align} 
The only nonzero terms are
\begin{align}
	L_0 &= 4\varepsilon_B \lambda m  = 4\lambda m \sqrt{m^2 +\frac{Q^2}{4 \sin^2 \frac{\theta_B}{2}}}\\
	L_1 &= 2 \lambda m q_B \cot \frac{\theta_B}{2} = 2\lambda mQ \cot \frac{\theta_B}{2} \, 
\end{align}
For the hadron tensor, the only relevant non-zero terms are 
\begin{align}
	W_0 &= 4G_M E_B q_B n_z +\frac{4(G_E -G_M)}{\tau+1} E_B q_B n_z\\
	W_1 &= 2G_M (4M^2 +Q^2) n_x 
\end{align}
Lastly, we multiply the components together and use Eq. (\ref{interference formula})
\begin{align}
	2\text{Re}[&\overline{\mathcal{M}_{1\gamma}\mathcal{M}_{1\pi}^*}]  = \frac{\pm 2 e^2\text{Re}[f_p^* g_p^*] }{Q^2(Q^2 +m_\pi^2)}\nonumber \\
	&\Biggl[8\lambda m\sqrt{1+\tau}QM\sqrt{m^2 +\frac{Q^2}{4} \csc^2(\theta_B /2)}F_1 n_z  \nonumber \\
	& + 4\lambda m Q\cot(\theta_B/2) G_M(4M^2 +Q^2)n_x\Biggr].
\end{align}

\bibliography{main.bib}

\end{document}